\documentclass{conm-p-l}

\usepackage[dvips]{graphics}

\theoremstyle{definition}

\theoremstyle{remark}

\numberwithin{equation}{section}

\newcommand{\lan}{\langle}
\newcommand{\ran}{\rangle}

\newcommand{\cp}{{\bf CP}}
\newcommand{\be}{\begin{equation}}
\newcommand{\ee}{\end{equation}}
\newcommand{\bq}{\begin{eqnarray}}
\newcommand{\eq}{\end{eqnarray}}

\newcommand{\no}{\nonumber\\}

\newcommand{\p}{\partial}
\newcommand{\la}{\lambda}

\newcommand{\al}{\alpha}

\newcommand{\U}{{\bf U}}

\begin{document}

\title{Quantum Computation by Geometrical Means}

\author{Jiannis Pachos}
\address{Institute for Scientific Interchange Foundation, 
Villa Gualino, Viale Settimio Severo 65, I-10133 Torino, Italy}
\email{pachos@isiosf.isi.it}
\thanks{The author was supported in part by TMR Network under the contract no. ERBFMRXCT96 - 0087.}

\date{January 19, 2000.}

\begin{abstract}
A geometrical approach to quantum computation is presented, where a non-abelian connection
is introduced in order to rewrite the evolution operator of an energy degenerate system
as a holonomic unitary. For a simple geometrical model we present an explicit construction 
of a universal set of gates, represented by holonomies acting on degenerate states.
\end{abstract}

\maketitle

\section{Prologue}

Abelian \cite{Sha} and non-abelian \cite{Wil} geometrical phases in quantum theory have been considered 
as a deep and fascinating subject. They provide a natural connection between the 
evolution of a physical system with degenerate structure and differential geometry. 
Here we shall present a model where these concepts can be explicitly applied
for quantum computation \cite{Zan}. 

The physical setup consists of an energy degenerate quantum system on which we perform an
adiabatic isospectral evolution described by closed paths in the 
parametric space of external variables. 
The corresponding evolution operators acting on the code-state 
in the degenerate eigenspace are given in terms of holonomies 
and we can use them as quantum logical gates. This is a generalization of the Berry phase 
or geometrical phase, to the non-abelian case, where a non-abelian adiabatic connection, 
$A$, is produced from the geometrical structure of the degenerate spaces.
In particular, on each point of the manifold of the external parameters there is a 
code-state attached and a
transformation between these bundles of codes is dictated by the connection $A$.

In order to apply this theoretical construction to a concrete example we employ
a model with $\cp^2$ geometry, that is a complex projective manifold with two complex 
coordinates. This is interpreted as a qubit \cite{Pac}. A further generalization with the tensor product of
$m$ $\cp^2$ models and additional interaction terms parametrized by the Grassmannian manifold, ${\bf G}(4,2)$,
is interpreted as a model of quantum computer.

The initial code-state is written on the degenerate eigenspace of 
the system. The geometrical evolution operator is a unitary 
acting on it and it is interpreted as a logical gate. 
Due to adiabaticity the geometrical part of the evolution operator 
has a dimensionality equal to the degree of degeneracy of the eigenspace. 
Specific logical gates given by holonomies are constructed for
a system with a tensor product structure resulting in
universality while at the end a quantum optical 
application is sketched.

\section{Coset Space Geometry}

A transformation $U(n)$ between the states $|\al \ran$, $\al=1,...,n$ can be 
realized by all possible sub-$U(2)$ transformations between any two 
of those  states, $|i\ran$
and $|j\ran$. A coset space can be produced as the factor with respect to 
some particular $U(2)$ symmetries of these transformations. 

Examples of such constructions are given in the following: 
\begin{itemize}
\item the $\cp^2$ projective space:
\begin{equation}
\cp^2\cong {U(3) \over U(2)\times U(1)}
\nonumber
\end{equation}
\end{itemize}
\begin{figure}[h]
\includegraphics{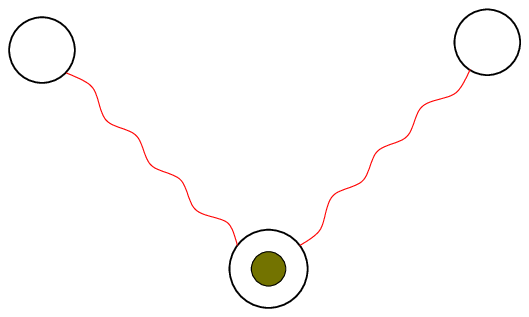}
\put(-240,165){$|1\rangle$}
\put(-110,165){$|2\rangle$}
\put(-175,100){$|\tilde2\rangle$}
\put(-210,130){$U_1(z_1)$}
\put(-90,130){$U_2(z_2)$}
\end{figure}
\vspace{-3cm}
The lines denote $U(2)$ transformations between the states represented here by ``holes''.
The $U(3)$ group could be interpreted by three lines connecting all the holes 
together.
The distinction between ``filled'' and ``unfilled'' holes is due to the coset structure, 
which factors out the symmetry transformations, between $|1\ran$ and $|2\ran$,
and denotes explicitly the non-symmetric ones between $|1\ran$ or $|2\ran$ and $|\tilde 2\ran$.
\begin{itemize}
\item the $\left(\cp^2 \right)^{\times m}\times \left({\bf G}(4,2)_{int}\right)^{\times (m-1)}$
product space:
\begin{equation}
\cdots{U(3) \over U(2)\times U(1)}\times {U(3) \over U(2)\times U(1)} \, ,\, \, \left. 
{U(4) \over U(2) \times U(2)} \right|_{int} \cdots
\nonumber
\end{equation}
\end{itemize}

\begin{figure}[h]
\includegraphics{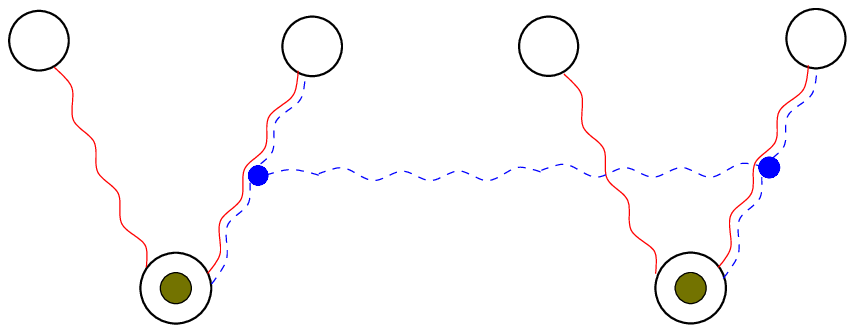}
\put(-360,130){$|1\ran$}
\put(-280,130){$|2\ran$}
\put(-320, 60){$|\tilde2\ran$}
\put(-210,130){$|3\ran$}
\put(-130,130){$|4\ran$}
\put(-170,60){$|\tilde 4\ran$}
\end{figure}
\vspace{-1cm}
Transformations can be performed between the states $\{|1\ran, |2\ran \}$ and 
$\{|3\ran, |4\ran \}$ due to their connections with the states $\{|\tilde 2\ran, |\tilde 4\ran\}$,
while an interaction in the tensor product space between $|24\ran$ and $|\tilde 2 \tilde 4 \ran$
gives transformations between those two sets. The transformations between only 
two states may be performed by linear operations with respect to $U(2)$ generators,
while the combined transformations between the two qubits can be produced by 
bilinear generators which act simultaneously on the states of both of the $\cp^2$ models.
The latter is denoted in the previous figure by the dashed lines, where the connection between 
the black dots indicates the simultanious action.

\section{Degeneracy, Adiabaticity and Holonomies}

Let us introduce the degenerate Hamiltonians $H_0^1$ and $H_0^m$ as follows
\begin{equation}
H_0^1= \left[  \begin{array}{ccccc}     
0 & 0 & 0 \\
0 & 0 & 0 \\
0 & 0 & 1 \\
\end{array} \right]\, ,\,\,\,\,\,\,\,\,\,\,\,\,
H_0^m= \sum_m H_0^1 \,\, .
\nonumber
\end{equation}
The orbit, that is the parametric manifold of the unitary transformations which 
preserves the degenerate spectrum of $H_0^1$ is given by $\cp^2$. 
A sub-manifold of the orbit of $H_0^m$ in which we are 
interested in is given by the 
$\left(\cp^2 \right)^{\times m}\times \left({\bf G}(4,2)_{int}\right)^{\times (m-1)}$
product manifold.

A general transformation parametrized by the $\cp^2$ space is given by
$\U ({\bf z}) := U_1(z_1) U_2(z_2)$, with $U_\al(z_\al)=\exp G_\al(z_\al)=\exp (z_\al 
|\al\ran\lan \tilde 2| -\bar z _\al | \tilde 2 \ran \lan \al |)$. The complex
parameter $z_\al$ may be decomposed as $z_\al=\theta_\al \exp i \phi_\al $. 
Due to the $2\times2$ sub-form of $G_\al(z_\al)$ we can rewrite $U_{\al}(z_\al)$ as 
\begin{equation}
U_{\al}(z_\al)={\bf 1}_{\al}^{\bot} + \cos \theta_\al {\bf 1}_{\al}+ \frac{ \sin \theta_\al }
{\theta_\al} G_\al(z_\al) \,\, ,
\nonumber
\end{equation}
where ${\bf 1}_\al^{\bot}={\bf 1} -{\bf 1}_\al$ and ${\bf 1}_\al=|\al\ran\lan \al|+|\tilde 2\ran \lan \tilde 2|$.
For the Grassmannian manifold ${\bf G}(4,2)$ we have, for example, 
the $U(2)$ rotation in the tensor product basis of two qubits,
between the states $|24\ran$ and $|\tilde 2 \tilde 4\ran$, given by 
$\U(z)=\exp (z |24\ran \lan\tilde2 \tilde 4| -\bar z|\tilde2 \tilde 4\ran\lan 24|)$, with $z=\theta \exp i\phi$.
The coordinates $\{\la^a\}=$\,\{{\boldmath{$\theta$}}, {\boldmath{$\phi$}}\} provide
the parametric space which the experimenters control.

In the four dimensional manifold $\cp^2$ with coordinates $\{\la^a\}$ a closed path, $C$, 
is drawn on a two-submanifold.
Consider this evolution to be adiabatic as well as isospectral which is provided by the formula 
$H(\la(t))={ \U}(\la(t)) H_0 { \U}^\dagger(\la(t))$.
As a result the state of the system, $|\psi(t)\ran$, stays on the same eigenvalue of the Hamiltonian, 
taken in our example to be $E_0=0$, without level-crossing. 

At the end of the loop $C$, spanned in time $T=N\Delta t$, when divided in $N$ equal 
time intervals, we obtain
\bq
|\psi(T)\ran&&={\bf T}e^{-i\int^T_0 {\U} H_0 { \U}^\dagger dt } \,\,\, |\psi(0)\ran
\no 
&&
={\bf T} \lim_{N\rightarrow \infty} \prod _{i=1}^N { \U}_i e^{-iH_0\Delta t}{ \U}_i^\dagger \,\,\, |\psi(0)\ran
\no 
&&
={\bf P} \lim_{N \rightarrow \infty}\left( {\bf 1} +\sum_{i=1}^N A_i\Delta \la_i\right) \,\,\, |\psi(0)\ran
\no 
\,\,\, \text{with} && \,\,\,\, A_i={ \U}_i^\dagger {\Delta { \U}_i \over \Delta \la_i} \,\,\,\,\, \text{and}
\,\,\,\,\, \U_i=\U(\la(t_i)) \,\, .
\nonumber
\eq
Hence, the state $|\psi(0)\ran$ acquires a geometrical unitary operator given by the holonomy 
of a connection $A$ as 
\begin{equation}
        \Gamma_A(C):= {\bf P} \exp \oint_C A \,\, ,\,\,\,\,\, \text{where}\,\,\,\,\,\,\,\,
        (A^{\la^a})_{\alpha \beta}:= \lan \alpha |{ \U}^\dagger(\la)
               {\p \over \p \la^a} { \U(\la)} |\beta\ran \,\, .
        \nonumber
\end{equation}
The states $|\alpha \ran$ and $|\beta \ran$ belong to the same degenerate eigenspace 
of $H_0$ and $\la^a$'s are the real control parameters.

The produced unitary operator is an effect of the non-commutativity of the control transformations
which produce effectively a curvature. In the case of the Berry phase produced
for example in front of the spin states of an electron when placed in a magnetic field, 
the non-commutativity is 
between the $U(2)$ control unitaries which change the direction of the magnetic field
in the three dimensional space. What is presented here is the generalization of the Berry phase 
to the non-abelian case.

The $\Gamma_A(C)$'s produced by $\cp^2$ for various loops $C$, generate the whole $U(2)$,
\begin{equation}
\{\Gamma_A(C) ;\,\,\, \forall \,\,\, C \in \cp^2\} \approx U(2) \,\, .
\nonumber
\end{equation}
In the case of $m$ qubits with their proper interactions, the produced group is $U(2^m)$,
\begin{equation}
\{\Gamma_A(C) ;\,\,\, \forall \,\,\, C \in 
\left(\cp^2 \right)^{\times m}\times \left({\bf G}(4,2)_{int}\right)^{\times (m-1)}\} 
\approx U(2^m) \,\, .
\nonumber
\end{equation}

\section{Quantum Computation}

In order to perform quantum computation by using the above constructions
we consider the following identifications:
\begin{center}

\vspace{0.3cm}

QUANTUM CODE $\equiv$ Degenerate States, $|\psi(0) \ran$ 

\vspace{0.4cm}

LOGICAL GATES $\equiv$ Holonomies, $\Gamma_A(C)$

\vspace{0.3cm}

      \end{center}
Let us first investigate the $\cp^2$ case. The basic question is
how we can generate a general $U(2)$ element by moving along a closed path, $C$.
Or in other words, for a specific $U\in U(2)$ which loop $C$ is such that $\Gamma_A(C)=U$.
In general, $\forall \,\,\, {\bf g} \,\,\, \in u(2) \,\,\, \exists 
\,\,\,\text{loop}\,\,\, C \in \cp^2$ manifold, such that $\Gamma_A(C)= \exp {\bf g}$,
which is the statement of irreducibility of the connection $A$ \cite{Zan}.
To answer the above question we perform the following analysis. The loop integral
\begin{equation}
\oint_C A=\oint_C A_{\la^1}d\la^1 +A_{\la^2} d \la^2+ \cdots
\nonumber
\end{equation}
is the main ingredient of the holonomy. Due to the path ordering symbol it is not
possible to just calculate it and evaluate its exponential, as in general the connection
components do not commute with each other. Still it is possible to 
consider the following restrictions in the position of the loop.
Choose $C$ such that: 
\begin{itemize}
\item it belongs to {\it one} plane $(\la^i, \la^j)=(\theta_i,\phi_j)$ or 
$(\theta_i, \theta_j)$, hence only two components of $A$ are involved,

\item the position of the plane is such that the connection, $A$, restricted on it become,
$A|_{(\la^i, \la^j)} = (A^{\la^i}=0,A^{\la^j})$, that is these two components commute with
each other. Still it is important that their related field strength component, 
$F_{ij}=\p_iA_j-\p_j A_i+[A_i,A_j]$, is non-vanishing in order to obtain a 
non-trivial holonomy. Such a requirement is possible for the $\cp^2$ model and for a wide class of 
other models.
\end{itemize}
On the planes where those conditions are satisfied the evaluation of the holonomy is trivially
given by just exponentiating the loop integral of the connection without worrying about 
the path ordering symbol.
Hence,
\begin{equation}
\Gamma_A(C)={\bf P} \exp \oint_C A = \exp (\Sigma {\bf  g}) = 
{\bf 1}_{2 \times 2} \cos \Sigma +{\bf g} \sin \Sigma \,\, ,
\nonumber
\end{equation}
where $\Sigma$ represents the area enclosed by the loop $C$ projected on the sphere
associated with the compactified $\cp^2$ manifold. This area
may be varied desirably. Furthermore, we are able to obtain a complete set of 
generators ${\bf g}$ by choosing $C$ to lie on different planes.
In detail we may obtain for ${\bf g}$ the following forms
\begin{eqnarray}
-i |\alpha \ran \lan \alpha| \,\,\,\,\,\,\,\,\,\,\,\,\,\, &:=& -i \sigma ^3_{\alpha} 
\,\, , \,\,\,  \al=1,\,\,2\,\,\,
\no 
-i(-i|1 \ran \lan 2| +i|2 \ran \lan 1 |)&:=& -i\sigma^2 \,\,, 
\no 
-i(|1 \ran \lan 2| + |2 \ran \lan 1 |)\,\,&:=& -i \sigma^1 \,\, . 
\nonumber
\end{eqnarray}
The $\sigma_\al^3$ generators are similar to a Berry phase and they are produced by 
paths $C_1$ on the $(\theta_\al, \phi_\al)$ planes. The 
corresponding holonomy is the exponential of this generator multiplied by 
the area, $\Sigma_1$, of the surface the path $C_1$ encloses when projected on a sphere 
$S^2(2\theta_\al,\phi_\al)$ with spherical coordinates $2\theta_\al$ and $\phi_\al$,
\begin{equation}
-i\sigma^3_\al :C_1 \in (\theta_\al,\phi_\al) \rightarrow \Gamma_A(C_1) =\exp -i \Sigma_1 \sigma_\al ^3 \,\, .
\nonumber
\end{equation}
The $\sigma^2$ generator is produced by a path $C_2$ along the plane $(\theta_1, 
\theta_2)$ positioned at $\phi_1=\phi_2=0$, 
while $\sigma ^1$ is produced by a path, $C_3$, along a parallel plane 
positioned at $\phi_1={\pi \over 2}$
and $\phi_2=0$. Their corresponding areas are $\Sigma_2$ and $\Sigma_3$. For example
\begin{equation}
-i\sigma^{2} :C_{2} \in \left.(\theta_1,\theta_2)\right|_{\phi_1=0, \phi_2=0} 
\rightarrow \Gamma_A(C_{2}) = \exp -i \Sigma_2 \sigma ^2 \,\, .
\nonumber
\end{equation}
Altogether we have $2^2$ independent generators spanning the 
Lie algebra of $U(2)$. 

For the case of the two qubit interaction the corresponding connection components are
given by 
\begin{equation}
A_{\theta}=\text{diag} (0,0,0,0)\,\,\, ,\,\,\, A_\phi=\text{diag} (0,0,0,-i\sin^2 \theta)
\nonumber
\end{equation}
which are written in the basis $\{|13\ran, |14\ran , |23\ran, |24\ran\}$.
A loop $C$ on the $(\theta,\phi)$ plane will produce the following holonomy
\begin{equation}
\Gamma_A(C)=\text{diag}(1,1,1,e^{-i\Sigma}) \,\, , \,\,\, \Sigma=\int_{D(C)} d\theta d\phi \sin 2 \theta \,\, ,
\nonumber 
\end{equation}
where $\Sigma$ can also be interpreted as an area on the sphere $S^2(2\theta,\phi)$.

\section{One and Two Qubit Logical Gates}

By performing appropriate loops we can obtain one qubit phase rotations as well as 
two qubit gates such as a controlled phase rotation $U_{CPH}$. 

Analytically, by spanning the indicated areas we may obtain
\begin{equation}
U_1=\exp \left[  \begin{array}{ccc}     
                 -i\Sigma_1 & 0 \\
                          0 & 0 \\
\end{array} \right] 
\,\,\, , \,\,\,\,
U_2=\exp \left[  \begin{array}{ccc}
                           0 & 0 \\
                           0 & i\Sigma_1\\
\end{array} \right]
\,\,\, , \,\,\,\,
U_3=\exp \left[  \begin{array}{ccc}     
                                  0 & -\Sigma_2 \\
                           \Sigma_2 & 0 \\
\end{array} \right] \,\, .
\nonumber
\end{equation}
The combinations
\begin{equation}
U_1 U_2=\exp (-i \sigma_3 \Sigma_1) 
\,\,\, , \,\,\, 
U_3=\exp (-i \sigma_2 \Sigma_2)
\nonumber
\end{equation}
can give any $U(2)$ transformation and hence any one qubit rotation.

For the two qubit gates we can construct easily the controlled rotation $U_{CPH}=\text{diag} 
(1,1,1,\exp -i \Sigma)$
between any pair of qubits. It is generated by a loop $C$ on the $(\theta,\phi)$ plane. 
Together with the one qubit rotations they provide a universal set of gates.

\section{Epilogue}

Apart from the intriguing theoretical formulation of holonomic computation 
there are several aspects of it, which have appealing technical advantages. 
Without overlooking the difficulties posed to an experimenter for 
performing continuous control over a system in order to span a loop, 
there are several unique characteristics of it, which await for exploitation.
For example, robustness  of the control procedure in terms of the spanned area,
according to errors in the actual form of the performed loop, as well as the isolation 
of the degenerate states as a calculational space may prove to be
advantages worth exploring.

In quantum optics displacing devices, squeezing devices and interferometers
acting on laser beams can provide the control
parameters for the holonomic computation. Each laser beam is placed in a non-linear Kerr medium 
with degenerate Hamiltonian $H_0=n(n-1)$, where $n$ is the photon numbering operator. 
The degenerate states $|0\ran$ and $|1\ran$ are the basis for encoding one qubit which is 
manipulated by displacing and squeezing devices. Any two qubit interactions can be implemented by
interferometers \cite{Cho}.

It is challenging for the experimenters to produced the desired closed paths.

\bibliographystyle{amsalpha}

\end{document}